\documentclass[aps,pra,twocolumn,superscriptaddress,groupedaddress]{revtex4}
\usepackage{amssymb}
\usepackage{dsfont}
\usepackage{color,graphicx} 
\usepackage{amsmath}

\usepackage{amsbsy} \usepackage{amsthm}
\usepackage{svg}
\usepackage{float} \usepackage{braket}
\usepackage{placeins}
\usepackage[colorlinks=true,citecolor=blue,linkcolor=red,urlcolor=red]{hyperref}
\usepackage{comment}
\usepackage{stackengine}

\usepackage{siunitx}
\newcommand{\dd}{\text{d}}
\newcommand{\xtilde}{{\raise.17ex\hbox{$\scriptstyle\sim$}}}
\providecommand{\norm}[1]{\lVert#1\rVert}

\providecommand{\ave}[1]{\langle #1 \rangle_t}
\providecommand{\x}{\mathbf{x}}

\title{Schr\"{o}dinger Newton}

\date{\today}

\begin{document}

\title{Enhancement of the effects due to the Schr\"odinger-Newton equation}

\author{Davide Giordano Ario Altamura}
\affiliation{Department of Physics, University of Trieste, Strada Costiera 11, 34151 Trieste, Italy}
\affiliation{Istituto Nazionale di Fisica Nucleare, Trieste Section, Via Valerio 2, 34127 Trieste, Italy}

\author{José Luis Gaona-Reyes}
\affiliation{Department of Physics, University of Trieste, Strada Costiera 11, 34151 Trieste, Italy}
\affiliation{Istituto Nazionale di Fisica Nucleare, Trieste Section, Via Valerio 2, 34127 Trieste, Italy}

\author{Elliot Simcox}
\affiliation{Department of Physics and Astronomy, University of Southampton,
Southampton SO17 1BJ, United Kingdom}

\author{Hendrik Ulbricht}
\affiliation{Department of Physics and Astronomy, University of Southampton,
Southampton SO17 1BJ, United Kingdom}

\author{Angelo Bassi}
\affiliation{Department of Physics, University of Trieste, Strada Costiera 11, 34151 Trieste, Italy}
\affiliation{Istituto Nazionale di Fisica Nucleare, Trieste Section, Via Valerio 2, 34127 Trieste, Italy}

\begin{abstract}

The Schrödinger-Newton (SN) equation introduces a nonlinear self-gravitational term to the standard Schr\"odinger equation, offering a paradigmatic model for semiclassical gravity. However, the small deviations it predicts from standard quantum mechanics pose significant experimental challenges. We propose a novel method to amplify such deviations through periodic modulation of the trapping frequency in a levitated mechanical oscillator. We identify specific regimes where the SN-induced effects on the dynamics of second moments are significantly enhanced—by up to six orders of magnitude compared to unmodulated setups. We show that this protocol remains feasible within current magnetic levitation technologies and enables distinguishability between standard and SN dynamics using measurable quantities such as the position variance. Our results pave the way for a viable experimental test of the SN equation, offering a new route to probe the interface between quantum mechanics and gravity.
\end{abstract}

\maketitle

{\it Introduction.--} General Relativity (GR) and Quantum Mechanics (QM) offer two fundamentally different descriptions of nature, the challenge of unifying them still remaining unresolved. While most approaches assume that gravity  must be quantized, the possibility that gravity remains  classical has also been considered~\cite{Møller1962,Rosenfeld1963,Diosi1984,diosi1989models,penrose1996gravity,carlip2008quantum,Kafri2014,Tilloy2016,Oppenheim2023}; within this less conventional framework, it has been suggested that, at the very fundamental level, the gravitational field is sourced by the expectation value of the stress-energy tensor describing quantized matter, rather than by the individual quantum states \cite{Møller1962,Rosenfeld1963,Bahrami2014}. In the weak field and non relativistic limit, one obtains the so-called Schrödinger-Newton (SN) equation \cite{Diosi1984}, a nonlinear modification of the Schrödinger equation incorporating a self-consistent gravitational potential generated by the system’s mass distribution as embodied by the wave function.

Let us denote by $\Psi_t(\x_1,...,\x_N)$ the wave function of a $N$-particle system. The SN equation reads:

\begin{equation}
\begin{split}
&i \hbar \partial_t \Psi_t(\x_1,...,\x_N)=(\hat{H}+\hat{H}_\text{grav})\Psi_t(\x_1,...,\x_N),
\end{split}
\end{equation}
where $\hat{H}$ is the Hamiltonian of the system encoding all  non-gravitational terms, and $\hat{H}_\text{grav}$ describes the gravitational interaction between the constituents of the system:
\begin{equation}
\hat{H}_\text{grav}=-G\sum_{j,k=1}^N m_jm_k \int \left\{ \prod_{l=1}^N \dd \x_l' \right\}\frac{|\Psi_t(\x_1',...,\x_N')|^2}{\norm{\x_j-\x_k'}}.
\end{equation}
This expression tells that the mass $m_j$  feels the gravitational potential of the mass $m_k$, where the mass density sourcing the corresponding gravitational field is  $m_k |\Psi_t(\x_1',...,\x_N')|^2$.
Due to this gravitational term, the SN is nonlinear in the wave function. 

Although the SN equation has a clear physical interpretation, it is not without problems, the most severe being that, if taken literally, it allows superluminal signaling~\cite{Bahrami2014}. This feature stems from its intrinsic nonlinearity rather than its non-relativistic nature. 

Interest in testing the SN equation has grown over the years \cite{VanMeter2011,Yang2013,Großardt2016,Gan2016,Colin2016,Großardt2017,helou2017measurable,Liu2023,Aguiar2024,Gruca2024,yan2025first}, as it represents one of the few counterexamples to a quantum theory of gravity.
Levitated mechanics is particularly suited~\cite{Yang2013,Giulini2014,helou2017measurable}, as it allows to reach optimal conditions for maximizing the gravitational effect, while keeping a high degree of control of the system. With suitable approximations and reducing the motion to one dimension, one arrives at the following equation  for the center-of-mass wave function $\psi_{t}(x)$~\cite{Yang2013}:
\begin{equation} \label{YangSN}
i \hbar \partial_t \psi_{t}(x)=\left(-\frac{\hbar^2}{2M}\partial^2_{x}+\frac{1}{2}M \omega^2 {x}^2+H_{\text{\tiny{SN}}} \right)\psi_{t}(x),    
\end{equation}
where $M$ is the total mass of the system, $\omega$ the  oscillation frequency, and the nonlinear Hamiltonian $\hat{H}_\text{\tiny{SN}}$ encodes the SN contribution:
\begin{equation}
H_\text{\tiny{SN}}=\frac{1}{2}M \omega_{\text{\tiny{SN}}}^2(x^2-2 x \ave{\hat{x}}+\ave{\hat{x}^2}).
\label{HSN}
\end{equation}
with $\ave{\hat{x}} = \langle \psi_t | \hat x | \psi_t \rangle$. The new frequency $\omega_{\text{\tiny{SN}}}$ depends on the mass distribution of the system; for a lattice whose mass is distributed in the vicinity of the lattice sites with a Gaussian distribution, the SN frequency $\omega_{\text{\tiny{SN}}}$ reads \cite{helou2017measurable}:
\begin{equation}
 \omega_{\text{\tiny{SN}}}= \sqrt{\frac{Gm}{6 \sqrt{\pi} \Delta x_{\text{zp}}^3}},
 \label{omegaSN}
\end{equation}
where $m$ is the atomic mass and $\Delta x_{\text{zp}}$ measures the fluctuations of their internal motion.  Eq.~\eqref{HSN} 
is valid as long as $\Delta x_{\text{zp}} > V_{\text{xx}}$, where $V_{\text{xx}}$ is the variance of the center-of-mass wave function. 

As for classical systems, the gravitational potential exerts no force on the center of mass (mean) motion~\cite{Yang2013,Giulini2014}:
\begin{equation} \label{firstmoments}
{\ave{\dot{\hat{x}}}}=\frac{\ave{\hat{p}}}{M}, \qquad {\ave{\dot{\hat{p}}}}=-M \omega^2 \ave{\hat{x}},
\end{equation}
while for the second moments, defined through the covariances $V_{\text{ab}}=\frac{1}{2}\ave{\{\hat{a},\hat{b}\}}-\ave{\hat{a}}\ave{\hat{b}}$, one has:
\begin{equation}
\label{secondmoments}
\begin{split}
\dot{V}_{\text{xx}} &=\frac{2}{M}V_{\text{xp}},\\
\dot{V}_{\text{xp}} &= \frac{1}{M}V_{\text{pp}}- M (\omega^2+\omega_{\text{\tiny{SN}}}^2)V_{\text{xx}},\\
\dot{V}_{\text{pp}} &=-2 M (\omega^2+\omega_{\text{\tiny{SN}}}^2)V_{\text{xp}}.
\end{split}    
\end{equation}
As noted in~\cite{Yang2013}, the  Hamiltonian $\hat{H}_\text{\tiny{SN}}$ modifies the rotation frequency  of the uncertainty ellipse: without this contribution, the ellipse rotates with frequency $\omega$, whereas with the inclusion of the SN contribution, it 
changes to $\omega_\text{q}~=~(\omega^2~+~\omega_{\text{\tiny{SN}}}^2)^{1/2}$. The experimental challenge is to discriminate the additional rotation induced by $\omega_{\text{\tiny SN}}$, competing with all realistic noise effects.

Experimental values of $\omega$ can be determined accurately for most levitated setups, with frequencies stable over the timescale of typical experiments and of sufficiently high quality factor~\cite{vinante2020ultralow, hofer2023high}. However, in order to predict $\omega$ accurately enough to discern $\omega_{\text{\tiny SN}}$ would require the experimental parameters to be known to a degree that is achievable, but beyond the level commonly used~\cite{westphal2021measurement, fuchs2024measuring}. For example, if $\omega_{\text{\tiny SN}}/\omega=1\%$ then experimental parameters such as particle's radius, mass and magnetization must all be known to this level to distinctively identify gravity as the cause for the observed effect. This is demanding and has not been achieved by experiments yet~\cite{yan2025first}.

\begin{figure}[h!]
    \centering
    \includegraphics[width=1\linewidth]{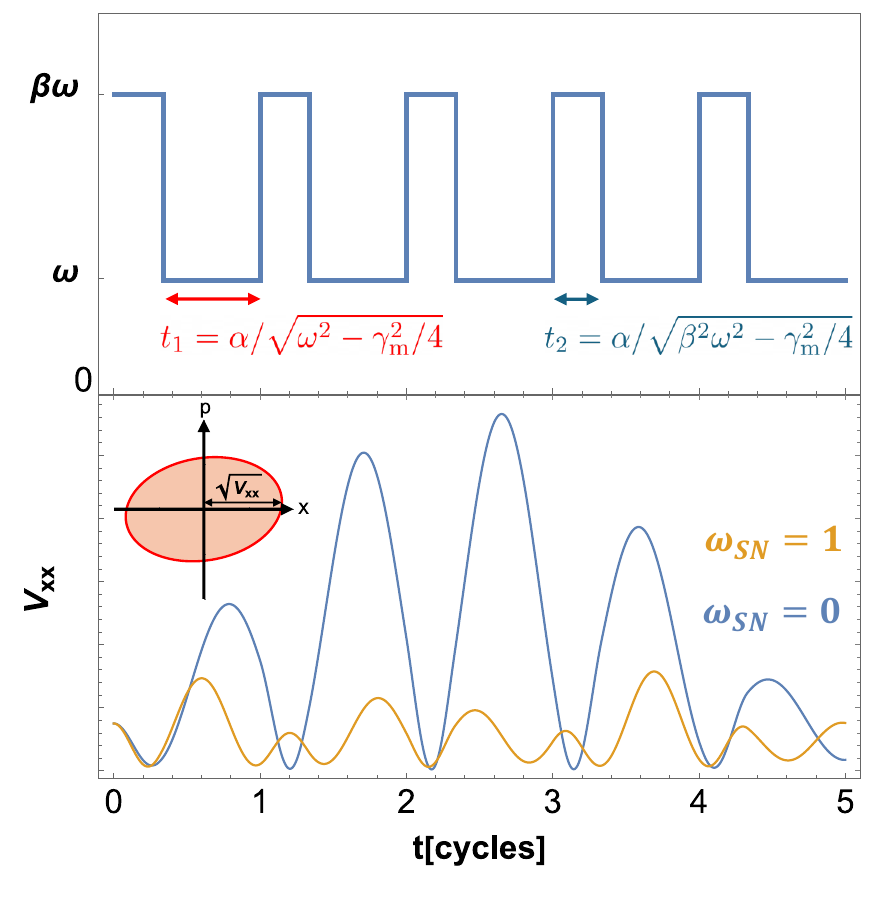}
    \caption{\textit{Top panel:} Modulation of the trapping frequency $\omega$.
    \textit{Bottom panel:} evolution of the position variance for a particular choice of $\alpha$ and $\beta$, with and without the SN term: the difference is amplified with a suitable modulation of $\omega$. (The inset shows the uncertainty ellipse and identifies $V_{\text{xx}}$.) }
    \label{Cartoon}
\end{figure}

In this work, we propose to enhance the deviations of the SN predictions with respect to standard quantum mechanics by suitably modulating the trap frequency $\omega$, in analogy with recent works where similar techniques have been employed in squeezing protocols \cite{wu2024squeezing}. The frequency modulation results in an appreciable difference of oscillation amplitude with and without the SN term; see Fig.~\ref{Cartoon}. We will show that this paradigm shift enhances significantly the chances for detection of the SN effect with today's technology: mechanical squeezing protocols based on frequency parametric driving~\cite{Pontin2014} or frequency jump operations~\cite{janszky1986squeezing,marti2024quantum} have been demonstrated; in levitated mechanical experiments, squeezing of thermal motional states has been shown~\cite{rashid2016experimental, duchavn2024experimental, kamba2025quantum} and the achieved squeezing factors on the level of 10~dB~\cite{muffato2025generation, wu2024squeezing, muffato2024coherent}.  

{\it Theoretical model.--} We replace the constant-frequency harmonic potential in Eq.~\eqref{YangSN} with a time-dependent periodic term $\hat{H}_\text{\tiny{Q}}=\frac{1}{2}M \omega^2_t x^2$, where the frequency $\omega_t$ reads
\begin{equation}
\omega_t= \begin{cases}\omega, & 0+n \tau \leq t<t_1+n \tau, \\ \beta\omega, & t_1+n \tau \leq t<(n+1) \tau,\end{cases}
\label{squarewave}
\end{equation}
with $t_1$, $t_2$, $\beta$ free parameters of the model, $\tau=t_1+t_2$, and $n$ a non-negative integer indicating the number of cycles of the evolution.

Although  the dynamics in Eq.~\eqref{YangSN} is nonlinear, and therefore not unitary, one can nevertheless construct an effective Heisenberg picture \cite{helou2017measurable}, and write down the corresponding equations of motion. In what follows, we will consider a more realistic scenario in which the oscillator is damped with decay rate $\gamma_\text{m}$ and subject to a classical thermal noise \cite{wu2024squeezing}. Straightforward calculations lead to the following expressions for the Heisenberg equations for the position and momentum operators:\cite{Yang2013,Colin2016}
\begin{equation}
    \begin{aligned}
\dot{\hat{x}}_t & =\hat{p}_t / M, \\
\dot{\hat{p}}_t & =-M( \omega^2_t\hat{x}_t+\omega^2_{\text{\tiny{SN}}}(\hat{x}_t-\ave{\hat{x}}))-\gamma_\text{m} \hat{p}_t+F_{\text {th,t }},
\label{Langevin}
\end{aligned}
\end{equation}
where $F_{\text {th,t}}$ is the thermal noise force with correlation: $\mathbb{E}[ F_{\text{th,t}}F_{\text{th,s}}]= 2 M \gamma_\text{m} k_\text{\tiny{B}} T \delta(t-s)$, with $k_\text{\tiny{B}}$ the Boltzmann constant and  $T$ the temperature of the bath, and $\mathbb{E}[\cdot]$ denotes the stochastic average over the noise. Without the thermal noise ($F_{\text{th,t}}=0$) and if $\omega_t$ is constant, the above Heisenberg equations, computed for the mean values of the position and momentum, reduce to Eq.~\eqref{firstmoments}.

The dynamical evolution of the average of the second moments ($V_{\text{ab}}=\frac{1}{2}\mathbb{E}[\langle\{\hat{a},\hat{b}\}\rangle_t]-\mathbb{E}[\ave{\hat{a}}]\mathbb{E}[\ave{\hat{b}}]$) \footnote{Note that now, differently from Eq.~\eqref{secondmoments}, an extra average $\mathbb{E}[\cdot]$ appears because the dynamics contains a noise term.} is given by 
\begin{equation} \label{Flononhom}
\dot{x}_t =P_t x_t + C,
\end{equation} 
with $x_t$ defined as $x_t^T~=~(V_{\text{xx}} \quad V_{\text{xp}} \quad V_{\text{pp}})$; the matrix $P_t$ reads:
\begin{equation}
    P_t=\begin{pmatrix}
    0 &\frac{2}{M}&0\\
-M \omega_{\text{q},t}^2 &-\gamma_\text{m} &\frac{1}{M}\\
0&-2 M \omega_{\text{q},t}^2&-2 \gamma_\text{m}
\end{pmatrix}, \label{Pmatrixdamp}
\end{equation}
where $\omega_{\text{q},t}~=~{(\omega^2_t+\omega^2_{\text{\tiny{SN}}})^{1/2}}$. The vector  $C$ is defined as

\begin{equation} \label{vectorC}
C^T=\left(0 \qquad M \omega_{\text{\tiny{SN}}}^2 F_{\text{xp},t} \qquad 2 M \gamma_\text{m}k_\text{B}T + 2 M \omega_{\text{\tiny{SN}}}^2 F_{\text{pp},t} \right),
\end{equation}
with $F_{\text{xp},t}$ and $F_{\text{pp},t}$ defined in Eq.~\eqref{Deviations}.  To simplify the analysis, we set the times $t_1$ and $t_2$ in Eq.~\eqref{squarewave} equal to $t_1=\alpha/\sqrt{\omega^2-\gamma_\text{m}^2/4}$ and $t_2=\alpha/\sqrt{\beta^2 \omega^2-\gamma_\text{m}^2/4}$, with the denominators  encoding the oscillation frequency in the presence of damping. Thus, we have restricted the number of free parameters to two \footnote{We notice that the choice $\alpha=\pi/2$ coincides with the setting in Ref.~\cite{wu2024squeezing}.}.

In what follows, we will consider two cases: first, the ideal one without damping  ($\gamma_\text{m}=0$, $F_{\text{th}}=0$), and then the more realistic one with damping ($\gamma_\text{m} \neq 0$). In both cases, since the matrix $P_t$ in  Eq.~\eqref{Flononhom} is periodic, with period $\tau$, we will resort to the Floquet-Lyapunov (FL) theory \cite{Gantmacher1959} (see Appendix \ref{FLtheory}) to characterize the stability of the second moments. The dynamical evolution in Eqs.~\eqref{Langevin} and Eq.~\eqref{Flononhom} is valid for an arbitrary quantum state, but we will restrict the analysis to Gaussian states, so that these equations suffice to characterize all higher moments of the system.

We analyze the evolution of the second moments of the system, where the SN term explicitly contributes to the dynamics; we will come back later on the evolution of the first moments, when describing the experimental implementation of the proposed protocol.

{\it Dynamics for $\gamma_\text{m}=0$.--} 
\begin{figure*}[ht!]
    \centering
    \includegraphics[width=1\linewidth]{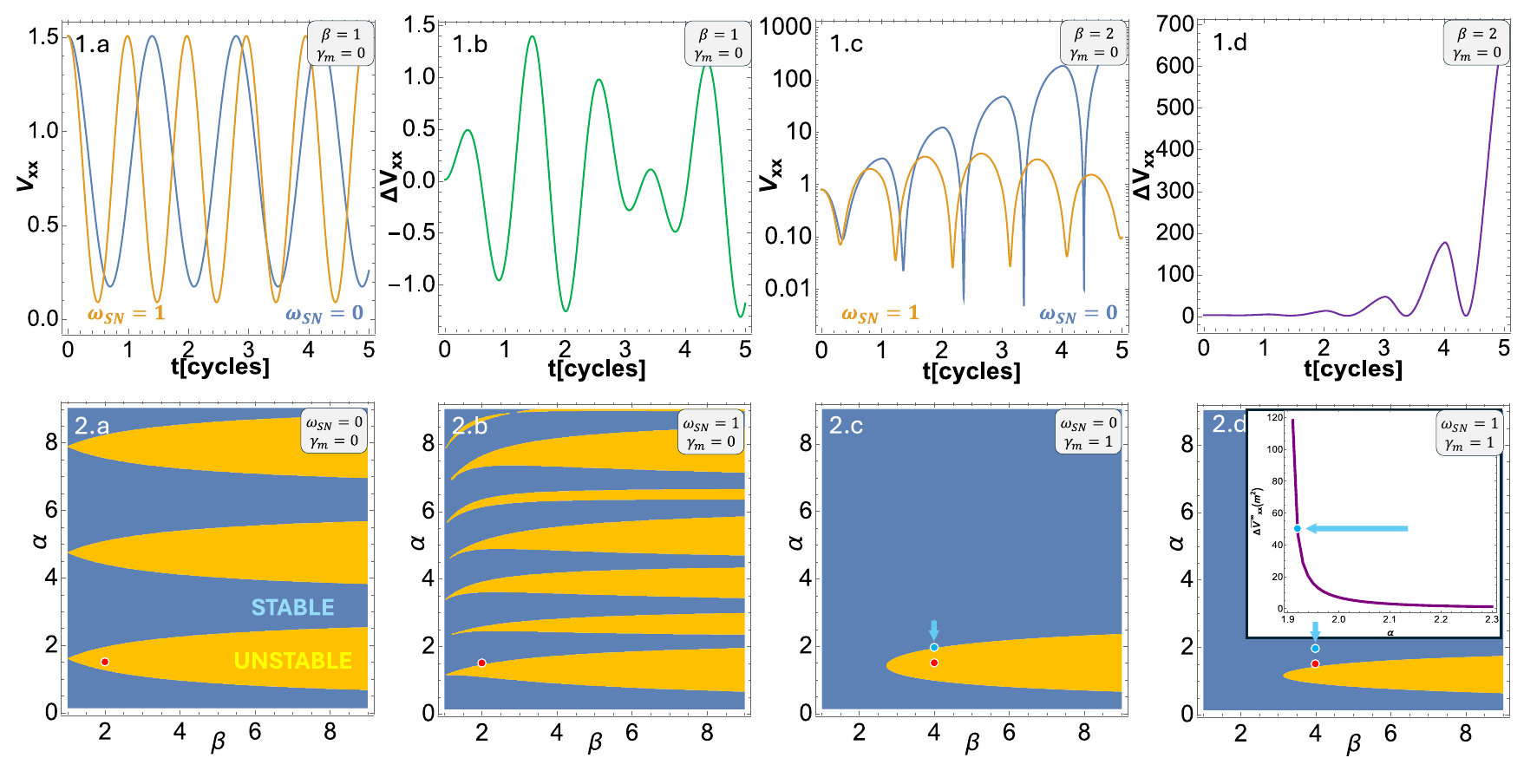}
    \caption{{\it Top panels.} Panel 1.a and 1.c show the time evolution of $V_{\text{xx}}$ for a particle of mass $M=1$ (in arbitrary units), without ($V_{\text{xx}}^0$, blue line) and with ($V_{\text{xx}}^\text{\tiny{SN}}$, yellow line) the SN term; we also set $\hbar=1$ and $\omega=1$. Panel 1.a refers to a constant frequency, and panel 1.c to a varying frequency, with $\beta=2$ and $\alpha=3/2$. 
    Panels 1.b and 1.d show the time evolution of the difference $\Delta V_{\text{xx}}=V_{\text{xx}}^0-V_{\text{xx}}^\text{\tiny{SN}}$ in the two cases with constant and varying frequency; in the second case, the difference is amplified. The plots are shown in units of the number $n$ of cycles, with $t=n \tau$ [cf. Eq.~\eqref{squarewave}]. {\it Bottom panels.} Panel 2.a and 2.b explore the stability and instability regions of the parameter space $(\alpha,\beta)$ without and with the SN term, but without damping. They differ in the two cases, offering the possibility to select values of $(\alpha,\beta)$ that make the dynamics stable in one case and unstable in the other, thus amplifying the SN effect; the red bullet is one such choice, corresponding to the parameters chosen for the plots in panels 1.c and 1.d.  
    Panels 2.c and 2.d show stability and instability regions with a damping term $\gamma_\text{m}=1$; again, they differ. Since the dynamics with both the SN term and  damping becomes non periodic [cf. Eqs.~\eqref{Flononhom}-\eqref{vectorC}], we have neglected the non periodic terms in identifying the stability and instability regions plotted in panel 2.d, as they give a negligible contribution. 
    The inset in panel 2.d shows the increase of the asymptotic difference $\Delta \bar{V}_{\text{xx}}^\infty = \bar V_{\text{xx}}^{\infty,0}-\bar {V}_{\text{xx}}^{\infty,\text{\tiny{SN}}}$ of the two envelopes of ${V}_{\text{xx}}$, without and with the SN term, when moving towards an instability region (calculations have been performed  including the non periodic terms). The blue bullet represents a choice of $(\alpha, \beta)$ amplifying $\Delta V_{\text{xx}}$ without generating an instability.
   }
    \label{Toymodel}
\end{figure*}
Let us first consider the case without damping, so that the vector $C$ vanishes and Eq.~\eqref{Flononhom} becomes homogeneous. For each cycle of the evolution of the system, the dynamics can be split in two parts with constant frequency. Let us denote by $P_1$ and $P_2$ the evolution matrices obtained by evaluating $P_t$ during the intervals $0 \leq t < t_1$ and $t_1 \leq t < \tau$ respectively. Then, the solution of Eq.~\eqref{Flononhom} after $n$ cycles is given by:
\begin{equation} \label{gamma0sol}
    {x}_n \equiv x_{n\tau}=L^n{x}_0, \qquad L=e^{P_2 t_2}e^{P_1 t_1},
\end{equation}
with $t_1=\alpha/\omega$ and $t_2=\alpha/(\beta \omega)$.

If $\beta = 1$, i.e.  $\omega_t$ is constant,  the rotation frequency  and the amplitude  of the uncertainty ellipse change if the SN term is present, as mentioned before. In terms of the evolution of the position variance $V_{\text{xx}}$, this change leads, for a given value of $\omega_{\text{\tiny{SN}}}$, to an increase of the oscillation frequency, besides a change of the amplitude, as shown in panels 1.a and 1.b of Fig.~\ref{Toymodel}. 

If $\beta \neq 1$, the modification in the amplitude of the oscillations of $V_{\text{xx}}$, with and without the SN term, can be customized; for specific choices of the parameters, there is an enhancement of the deviations of the SN dynamics with respect to standard quantum one; see panels 1.c and 1.d of Fig.~\ref{Toymodel} for an example of this kind.

To assess when such an amplification occurs, one can study the asymptotic behavior of $V_{\text{xx}}$ with and without the SN term, by 
examining the stability of the solution in Eq.~\eqref{gamma0sol}, which can be characterized by using the FL theory \cite{Gantmacher1959}. The solution is stable if all the eigenvalues of $L$ are smaller or equal than 1 in modulus. This is shown in panels 2.a and 2.b of Fig.~\ref{Toymodel} for the case under study; as anticipated, there are regions where the solution is stable without the SN term in the dynamics, but is unstable with the SN term, and vice versa. 

In doing so, one has to keep in mind that the condition $V_{\text{xx}} < \Delta x_{\text{zp}}$ must always be  satisfied for Eq.~\eqref{HSN} to hold. There are two possibilities for achieving this: 
one is to remain within a stable regime with and without the SN term, but near a zone of instability to enhance the SN effect without making it diverge. Alternatively, one can  choose parameters corresponding to an unstable solution, for example only when the SN term is absent, but stable when the SN term is present, and stop the time evolution when the condition $V_{\text{xx}} < \Delta x_{\text{zp}}$ is no longer fulfilled. We will come back on these strategies later.

{\it Dynamics for $\gamma_\text{m} \neq 0$.--} 
Let us now consider the effect of a non-vanishing damping rate. The FL analysis of the stability of the solutions can still be carried out, but now the vector $C$ in Eq.~\eqref{vectorC} no longer vanishes and therefore the system in Eq.~\eqref{Flononhom} is not homogeneous. The full analysis is presented in Appendix~\ref{FLtheory}. The dynamics still allows for stable and unstable solutions, which differ with and without the SN term and also with respect to the case without damping, as shown in panels 2.c and 2.d of Fig.~\ref{Toymodel}. 

Also in this case, one can  adopt the strategy of carefully choosing the parameters $(\alpha,\beta)$ to enhance the SN effect.
The components of $x_t$ oscillate over time; as before, we focus on the first component ${V}_{\text{xx}}$. Let us consider the envelope 
$\bar{V}_{\text{xx}}$ constructed by interpolating the maxima of the oscillations of $V_{xx}$. In the inset of panel 2.d in Fig.~\ref{Toymodel}, we plot the asymptotic difference $\Delta \bar{V}_{\text{xx}}^\infty = \bar V_{\text{xx}}^{\infty,0}-\bar {V}_{\text{xx}}^{\infty,\text{\tiny{SN}}}$ of the position variance without and with the SN term, for a fixed value of $\beta$, and letting $\alpha$ vary near an instability region. We see that as $\alpha$ approaches an instability region $\Delta \bar {V}_{\text{xx}}^\infty$ becomes larger, and thus the effect of the SN term becomes more relevant. 

\begin{table}[t]
    \centering
    \begin{tabular}{|c|c|c|c|c|}
    \hline
    $T(\mathrm{K})$&$V_{\text{xx}}(t=0)$&$V_{pp}(t=0)$&$V_{\text{xp}}(t=0)$ & $\Delta x_{\text{zp}}(\mathrm{m})$ \\
       \hline
       $10$& $\hbar/(2 M\omega)$&$\hbar M\omega/2 $ &0& $3.5\times10^{-12}$\\
       \hline  \hline
       $M(\mathrm{kg})$ &  $\omega (\mathrm{Hz})$ & $\omega_{\text{\tiny{SN}}}(\mathrm{Hz})$ & $m(\mathrm{kg})$ &$\gamma_\text{m}(\mathrm{Hz})$\\
        \hline
       $10^{-5}$& $5\times 2\pi$& $1.2\times10^{-1}$ & $9.3\times10^{-26}$& $10^{-1}$ \\
       \hline
    \end{tabular}
    \caption{List of the experimental parameters considered in the analysis. We further set $\beta=2$ and $\alpha=1.911$ for computing $\Delta \bar V_{\text{xx}}$ as reported in Fig. \ref{SecondMomentsAmplitude} and $\beta=2$ and  $\alpha=1.910625$ for $\Delta \bar V_{\text{xx}}$ reported in Fig. \ref{SecondMomentsAmplitudeUnstable}.}
    \label{RealParameters}
\end{table}

{\it Enhancement of the SN deviations.--} Having understood the dynamics of the system, we can now move to consider a realistic experimental situation. 
In what follows, we consider the parameters in Table~\ref{RealParameters}, which are compatible with levitated mechanical systems \cite{gonzalez2021levitodynamics}.

In addition to the condition $V_{\text{xx}}<\Delta x_{zp}$, the system  must remain confined in the trap during the experiment; to this end, one has to control the behavior of the first moments. Contrary to the second moments,  the stability regions of the first moments are independent of  $\omega_{\text{\text{SN}}}$; see Appendix ~\ref{AppFirstMoments} for further details. 
Then, to observe the signature of the SN term, one can either work in a stable regime for both the first and second moments, or one can choose values of $\alpha$ and $\beta$ such that one of the two cases (either with $\omega_{\text{\tiny{SN}}}=0$ or $\omega_{\text{\tiny{SN}}}\neq 0$) corresponds to an unstable region, and the other to a stable one. We discuss both cases.

\begin{figure}[t!]
    \centering
    \includegraphics[width=1\linewidth]{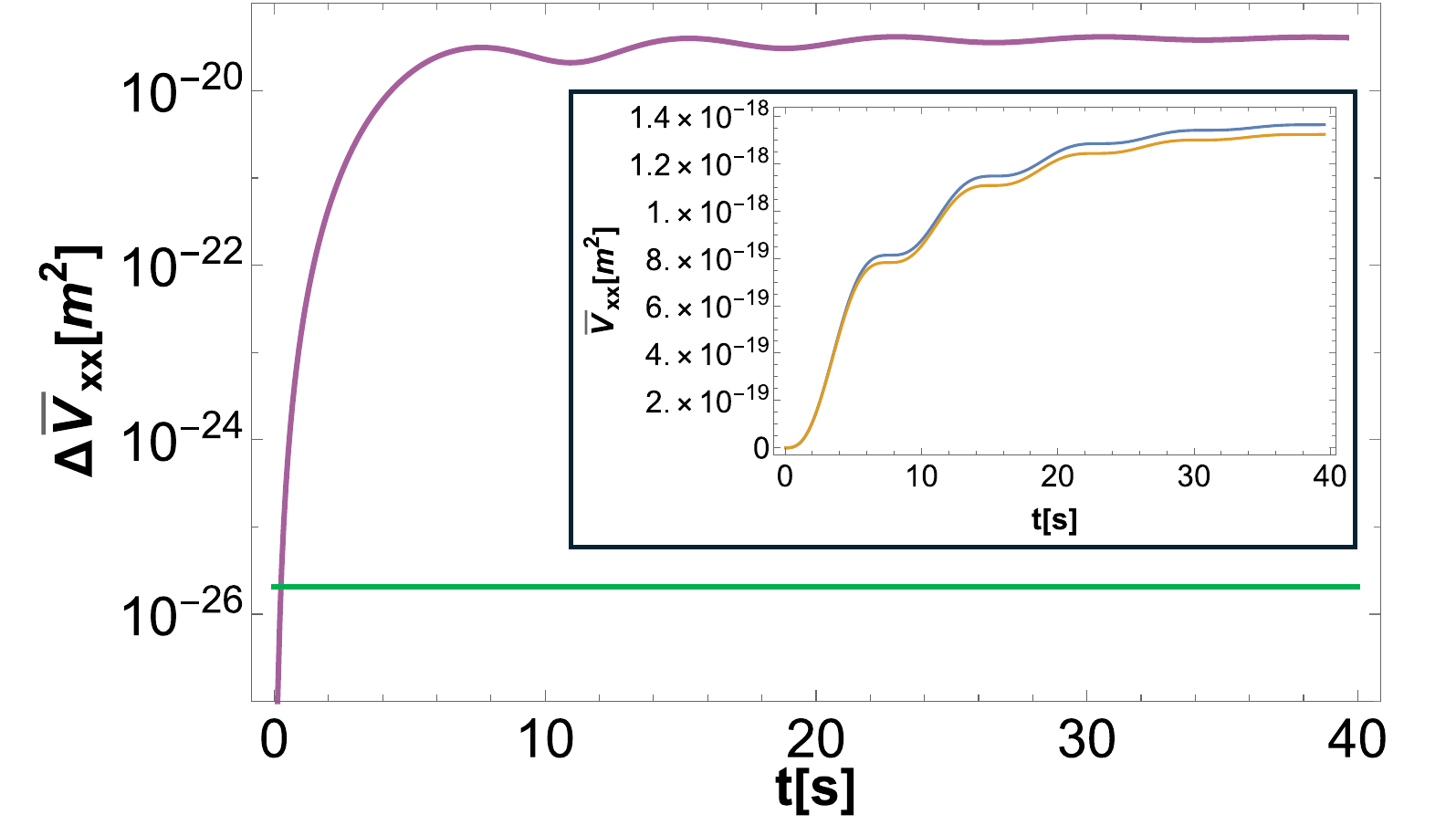}
    \caption{The purple line shows the time evolution  of the difference $\Delta \bar V_{\text{xx}}= \bar V_{\text{xx}}^0-\bar V_{\text{xx}}^ {\text{\tiny{SN}}}$ of the upper envelope of $V_{xx}$, without ($\bar{V}_{\text{xx}}^0$) and with ($\bar{V}_{\text{xx}}^{\text{SN}}$) the SN term, for a time-dependent frequency $\omega_t$, with $\beta=2$ and $\alpha = 1.911$. The inset shows the time evolution of $\bar{V}_{xx}^0$ (blue line) and $\bar{V}_{\text{xx}}^{\text{SN}}$ (yellow line) separately. The green line represents the maximum value that $\Delta \bar V_{\text{xx}}$ can attain when $\beta=1$ (no frequency modulation). The solution of Eq.~\eqref{Flononhom} is stable in both cases.  The other parameters are chosen as in Tab.~\ref{RealParameters}.}
    \label{SecondMomentsAmplitude}
\end{figure}

In Fig.~~\eqref{SecondMomentsAmplitude}, we show the time evolution of $\Delta \bar V_{\text{xx}}= \bar V_{\text{xx}}^0-\bar V_{\text{xx}}^ {\text{\tiny{SN}}}$, with a frequency modulation $\beta = 2$ (purple line) and compare it with the maximum value that $\Delta \bar V_{\text{xx}}$ reaches without any modulation (green line). We note that the system's initial state, presented in Tab.~\ref{RealParameters}, is assumed to be pure~\footnote{Since the Schr\"odinger-Newton equation is nonlinear, there is no clear way how to treat density matrices, because in general they can correspond to different ensembles of pure states, which evolve differently under this SN dynamics. To avoid this, we consider only initially pure states.}.
We see that $\Delta \bar V_{\text{xx}}$ increases over time, and is enhanced by six orders of magnitude with respect to the case with no frequency modulation. In the inset of Fig.~\eqref{SecondMomentsAmplitude} we show separately the time evolution of both $\bar V_{xx}^0$ and $\bar V_{xx}^{\text{\tiny{SN}}}$.  

\begin{figure}[t!]
    \centering
    \includegraphics[width=1\linewidth]{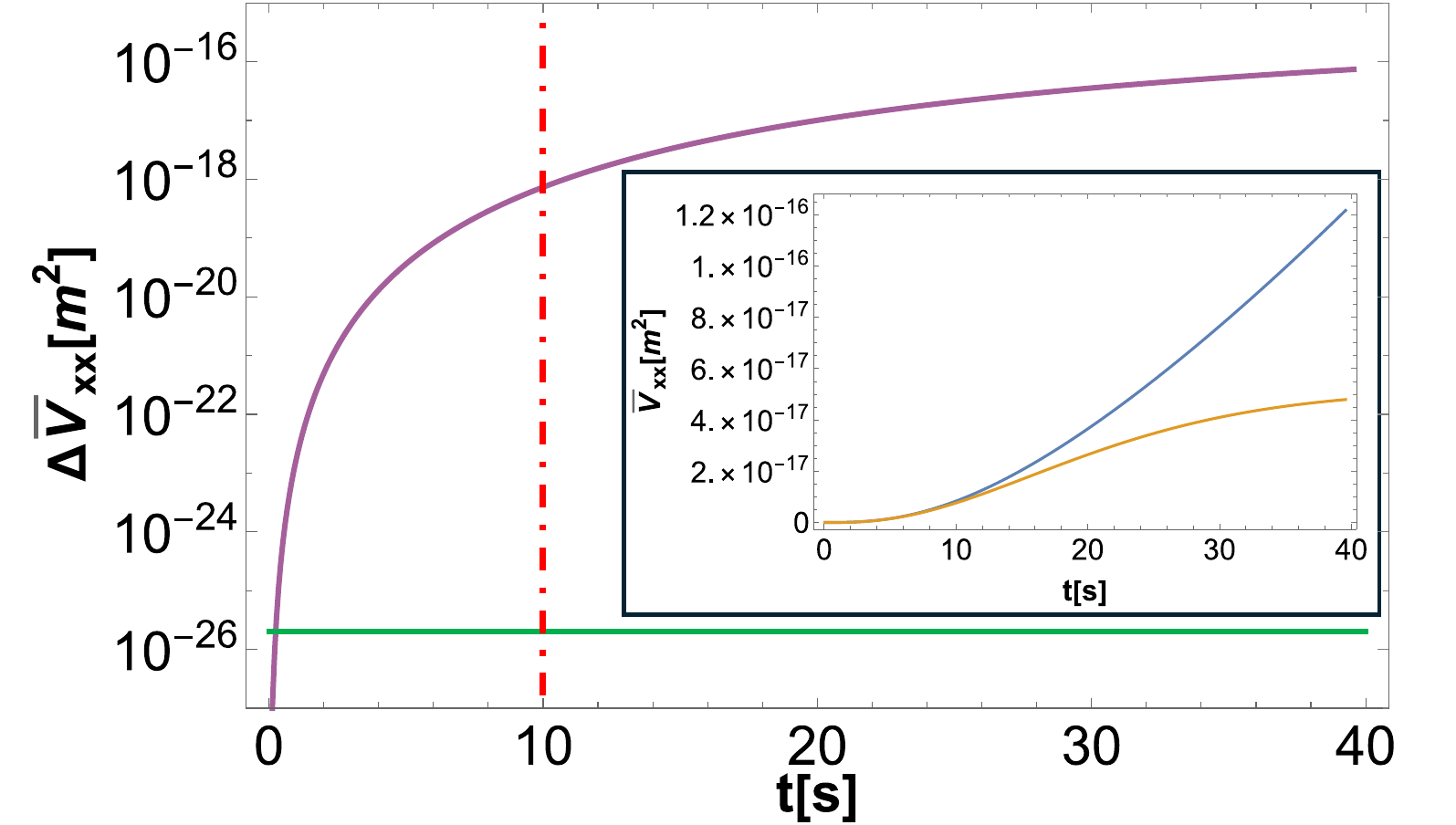}
    \caption{
    Same as in Fig.~\ref{SecondMomentsAmplitude}, but with $\alpha =1.910625$. Now, the solution of Eq.~\eqref{Flononhom} is unstable for $\omega_{\text{SN}}=0$ and stable for $\omega_{\text{SN}} \neq 0$. The red dash-dotted line indicates the time when the particle leaves the trap ($|\langle \hat{x}\rangle|>10^{-3} \mathrm{m}$).}
\label{SecondMomentsAmplitudeUnstable}
\end{figure}

The other strategy to amplify the SN effect is to move to an instability zone. This is considered in Fig.~\eqref{SecondMomentsAmplitudeUnstable}, where the parameters $(\alpha,\beta)$ are chosen in such a way that the solution is unstable for $\omega_{\text{\tiny{SN}}}=0$, but stable for $\omega_{\text{\tiny{SN}}} \neq 0$. We see in the corresponding inset that $\bar V_{\text{xx}}^0$ grows exponentially, while $\bar V_{\text{xx}}^{\text{\tiny{SN}}}$ eventually stabilizes. Notice that the system eventually escapes from the trap. Assuming a trap of width $10^{-3}\,\mathrm{m}$  leads to a time of confinement in the trap of $t \simeq 10\,\mathrm{s}$, as detailed in Appendix~\ref{AppFirstMoments}; both values are compatible with current experimental platforms. 

{\it Experimental feasibility.--} The protocol for the experimental realization consists in first preparing a pure initial motional state as close as possible to the quantum ground state, and then in applying the frequency modulation as illustrated in~Fig.~\ref{Cartoon}. Monitoring the amplitude increase will allow to discern dynamics with and without SN effects. 

We consider levitated mechanical systems which are particularly relevant for gravity experiments in the quantum regime~\cite{bose2025massive} and which have seen rapid experimental progress in the past decade of its different variants~\cite{gonzalez2021levitodynamics}. Here we focus on magnetic levitation, which offers lowest noise conditions and high mechanical quality factors and passive Meissner traps~\cite{Timberlake2020} as well as active coil traps of superconductors exist~\cite{hofer2023high}. Parameters for a magnetic Meissner trap of a millimeter-sized ferromagnet (Nd$_2$Fe$_{14}$B) above a type-1 superconducting and with quantum-limited position detection by SQUID~\cite{vinante2019testing} are summarized in Tab.~\eqref{RealParameters} and results are shown in Figs.~\ref{SecondMomentsAmplitude} and \ref{SecondMomentsAmplitudeUnstable}. Experiments are performed at low temperature ($<10$~K) which significantly reduced all thermal noise effect, at ultra-high vacuum ($<10^{-9}$~hPa) which reduces gas collisional noise, and with advanced vibration isolation to reduce heating from seismic and mechanical excitation~\cite{fuchs2024measuring}, the trap is passive so that no noise is introduced by the trapping interaction to affect the motion of the particle~\cite{vinante2020ultralow}.

The more challenging part of the experiment is to prepare an initial pure state, such as a number (Fock) state and in particular the mechanical quantum ground state. In magnetic levitation, initial steps for cooling towards the ground state have been taken. The variance of a harmonic system driven by thermal noise at temperature $T$ is~\cite{Timberlake2020}: $\bar V_{\text{xx}}= {(k_B T)}/{(M \omega^2)}$, where operation at the thermal noise limit has been achieved at $T<10$~K~\cite{vinante2020ultralow} and therefore all noises, including effects of mechanical vibration can be controlled to the needed levels~\cite{fuchs2024measuring}. The motional mode temperature has been cooled by feedback to the environmental temperature $T<1$~K~\cite{Timberlake2024} which already allows one to measure variance at signal-to-noise ratio (SNR) of 100. It is expected that the quantum ground state will be reached in low-frequency magneto-levitation in the next five years or so by feedback cooling~\cite{hofer2023high}.

A fundamental limit for increasing the oscillation amplitude is given by trap loss of the levitated particle. The size of the stable trap is about 
one millimeter (or even larger) for the magnetic trap. We estimate that trap loss would happen well after the clear separation between the dynamics with and without SN effects becomes apparent, and for more than 6 orders of magnitude amplitude amplification starting from the spatial extension of the zero-point motion at ground state. Therefore, we conclude that our proposal is feasible with today's levitated mechanics technology.

{\it Discussion and Outlook.--}
Due to the weakness of gravity, the deviations predicted by the SN equation with respect to standard QM are very weak, therefore their experimental verification poses a challenge. Here we have shown that introducing a time-dependent frequency in the Hamiltonian $\hat{H}_\text{\tiny{Q}}$ [cf. Eq.~\eqref{squarewave}] leads to an enhancement of the  SN effect. Through the use of the FL theory, we have characterized the behavior of the solutions and identified the regions of parameter space where the deviations predicted by the SN terms are maximized.

We have studied the experimental realization with levitated mechanics and found feasibility of the proposal with current technology. Our frequency modulation protocol enhances the ability to discern SN effects by conditional amplification of the the amplitude of motion of a massive harmonic oscillator. 

Investigating the stability regions in Fig.~\eqref{SecondMomentsAmplitude} created by the SN term may become impractical in systems where the range of $\beta$ is limited. However, one of the active areas of research in the levitated community is the increase of quality factor by reducing damping $\gamma_m$, which would allow smaller values of $\beta$ to be used, and hence allow access to larger unstable regions.

A further method of detecting the SN effect would be to compare several particles with the same expected trap frequencies but differing mass density of various ferromagnetic materials. For magnetic particles, $\omega_{\text{\tiny{SN}}}$ does not depend on the magnetization, but the trapping frequency does depend on it~\cite{Headley2025}. Hence it is possible to increase the mass density and the magnetization to keep $\omega$ constant while increasing $\omega_{\text{\tiny{SN}}}$. This would require specific care to quantify the trap's dimensions, the particles density and the particles magnetization to the degree that the increase in $\omega_{\text{\tiny{SN}}}$ can be identified. 
As our protocol is  mass independent, other experimental platforms may be employed. We considered optical levitation of silica. 
This platform 
turns out to be less promising than the magnetic levitated one: simulations show that the dynamics is in the regime $\Delta x_{\text{zp}} \le V_{\text{xx}}$ in which the SN effect becomes  smaller \cite{yan2025first,giulini2013gravitationally}, and the approximations on which this work is based do not hold anymore.
Other mechanical platforms, especially low frequency ones, such as clamped optomechanics, LIGO or torsion pendulums~\cite{aspelmeyer2014cavity}, might be valid experimental alternatives.

{\it Acknowledgments .--}
D.G.A.A., J.L.G.R. and A.B. acknowledge support from the EU EIC Pathfinder project QuCoM (101046973), INFN and the University of Trieste. A.B. acknowledges further support from  the PNRR MUR projects PE0000023-NQSTI. E.S. and H.U. acknowledge funding from the EU Horizon Europe EIC Pathfinder project QuCoM (10032223), from the UK funding agency EPSRC (grants EP/W007444/1, EP/V035975/1, EP/V000624/1), and from the Leverhulme Trust (RPG-2022-57).

\bibliography{biblio}

\clearpage
\appendix

\section{Floquet-Lyapunov theory}\label{FLtheory}
In what follows, we review the main elements of the Floquet-Lyapunov (FL) theory; we refers to Refs.~\cite{Gantmacher1959,chicone2006ordinary,slane2011analysis} for an extensive discussion. 

{\it Lyapunov criterion for the stability of solutions.--} 
Let us consider a vector $x_t=(x_1 \quad x_2 \quad \cdots \quad x_n)^T$, whose $n$ components vary over time. Suppose that $x_t$ satisfies the first-order differential equation $\dot{x}_t=P_t x_t$, where $P_t$ is a time-dependent square matrix of dimension $n$. An {\it integral matrix} $X_t$ is a square matrix whose columns are $n$ linearly independent solutions of the above system. As such, also $X_t$ satisfies the differential equation
\begin{equation}\label{intmat}
\dot{X}_t=P_t X_t.
\end{equation}
In particular, if the matrix $P_t$ is periodic with period $\tau$,  one can show that the matrix $X_{t+\tau}$ is again solution of Eq.~\eqref{intmat}; therefore $X_{t+\tau}=X_t V$, with $V$ a constant non-singular matrix. Moreover, the matrix $L_t=X_{\tau}e^{-\frac{t}{\tau}\ln V}$ is also periodic, with period $\tau$, and is a Lyapunov matrix \cite{Gantmacher1959}. 

Through a Lyapunov transformation, one can recast the original system in Eq.~\eqref{intmat} into one with constant coefficients, and as such the system is said to be reducible.  Specifically, by defining the matrix $Y_t=L_tX_t$, one can show that $Y_t$ satisfies 
\begin{equation} \label{redsist}
    \dot{Y}_t=A Y_t, \qquad A=\frac{1}{\tau}\ln V.
\end{equation}
The eigenvalues $\{a_j\}_{j=1}^n$ of $A$ are related to the eigenvalues $\{v_j \}_{j=1}^n$ of $V$ by the relation $a_j=(1/\tau) \ln v_j$, $j=1,...,n$. 

Lyapunov's criterion establishes that the solution of the system $\dot{x}_t=P_tx_t$ is stable if all the eigenvalues $\{v_j \}_{j=1}^n$ of $V$ satisfy $|v_j|\le1$, and unstable if at least one of them is such that $|v_j|>1$.

{\it Solution of an nonhomogeneous system with periodic coefficients.--} 
Let us suppose, without loss of generality, that $X_t$ is the {\it fundamental matrix}, which is the solution of the system in Eq.~\eqref{intmat} with the initial condition $X_0=I$. (Given an integral matrix $\Psi_t$, the solution can always be rescaled to obtain the fundamental matrix solution $X_t=\Psi_t\Psi_0^{-1}$). The solution of the system
\begin{equation}
\dot{x}_t=P_t x_t + C_t,
\end{equation}
where $C_t$ is a time dependent $n$-dimensional vector, is~\cite{chicone2006ordinary}
\begin{equation} \label{varpar}
x_t=X_tx_0+X_t\int_0^t X^{-1}_s C_s \dd s  .   
\end{equation}
After $n$ periods, the solution can be written as
\begin{equation} \label{solnperiods}
x_{n \tau}=X_{\tau}^n x_0 + \sum_{j=1}^n X_{\tau}^j \int_0^{\tau} X_s^{-1}C_s\dd s.
\end{equation}

The analysis in Ref.~\cite{slane2011analysis} shows that the term $X_\tau^n$ in Eq.~\eqref{solnperiods} satisfies $\lim_{n\rightarrow \infty}X_{\tau}^n =0$ if all eigenvalues $\{v_j\}_{j=1}^n$ of the matrix $X_\tau$ are such that  $|v_j|<1$; it converges to a finite (but not necessarily zero) matrix if the largest eigenvalue is 1 in modulus, is the only eigenvalue in the unit circle, and is semisimple (i.e., its algebraic and geometric multiplicities are the same). If there are other eigenvalues different from 1 in the unit circle, but they are not semisimple, $X^n_{\tau}$ remains bounded, but is non-convergent. In all other cases, $X^n_{\tau}$ diverges. 

The integral in the second term of Eq.~\eqref{solnperiods} yields a  matrix independent of $n$, therefore to asses the long time behavior of the solution, one needs to examine only the sum $\sum_{j=1}^n X_{\tau}^j$. The Neumann series, defined by $\sum_{j=0}^\infty X_{\tau}^j$ converges to $(I-X_{\tau})^{-1}$ if and only if all the eigenvalues $\{v_j \}_{j=1}^n$ of the matrix $X_\tau$ satisfy $|v_j|<1$, or equivalently, if $\lim_{n\rightarrow \infty} X^n_{\tau}=0$. In this case, the sum in Eq.~\eqref{solnperiods} converges to $(I-X_{\tau})^{-1}X_{\tau}$.

Summing up all elements, one has that the convergence of the solution in Eq.~\eqref{solnperiods} is guaranteed as long as all the eigenvalues $\{ v_j \}_{j=1}^n$ of the matrix $X_\tau$ satisfy $|v_j|<1$. All other cases need a closer examination, as the solution might be either bounded but non-convergent, or divergent.

\section{Time Evolution of the First  Moments}
\label{AppFirstMoments}

We study the dynamical evolution of the vector $y_t^T=(\mathbb{E}[\ave{\hat x}] \quad \mathbb{E}[\ave{\hat p}])$, where $\hat{x}$ and $\hat{p}$ satisfy Eq.~\eqref{Langevin}:
\begin{equation}
\mathbb{E}[\ave{\dot{\hat{x}}}]=\frac{1}{M}\mathbb{E}[\ave{\hat{p}}], \quad \mathbb{E}[\ave{\dot{\hat{p}}}]=-M \omega_t^2 \mathbb{E}[\ave{\hat{x}}]-\gamma_m \mathbb{E}[\ave{\hat{p}}].
\label{EqFirstMoments}
\end{equation}
Since the dynamics is homogeneous, from Eq.~\eqref{varpar}, the solution reads
\begin{equation} \label{solfirstmom}
    y_{n\tau}=A^n y_0, \qquad A=e^{A_2 t_2}e^{A_1 t_1},
\end{equation}
where the matrices $A_1$ and $A_2$ are given by
\begin{equation}
    A_1=\begin{pmatrix}
    0&1/M\\
    -M\omega&-\gamma_\text{m}
\end{pmatrix},\quad A_2=\begin{pmatrix}
   0&1/M\\
    -M\beta\omega&-\gamma_\text{m}
\end{pmatrix}.
\end{equation}
We can apply the FL theory to analyze the stability of the solution in Eq.~\eqref{solfirstmom}. In particular, when all the eigenvalues of $A$ are smaller than 1 in modulus, the first moments are stable, and thus the oscillatory motion remains bounded over time.

The eigenvalues can be computed explicitly, but their  expression is rather long and of little use; we report it only for the case $\gamma_m=0$:
\begin{eqnarray}
    \lambda_1 & =& [-2 \left(\beta ^2+1\right) \sin ^2(\alpha )+4 \beta  \cos ^2(\alpha )+(\beta +1) \nonumber \\
    &  &\sqrt{-2 (\beta +1)^2 \cos (2 \alpha )+2 (\beta -6) \beta +2} | \sin (\alpha )| ]/4 \beta \nonumber, \\
    \lambda_2 &= &-[-(\beta +1)^2 \cos (2 \alpha )+(\beta +1) \nonumber \\
    & & \sqrt{-2 (\beta +1)^2 \cos (2 \alpha )+2 (\beta -6) \beta +2} | \sin (\alpha )| \nonumber \\
    & & +(\beta -1)^2]/4 \beta .
\end{eqnarray}
In this case, the eigenvalues do not depend on $M$ and $\omega$ but only on the free parameters $\alpha$ and $\beta$ of the protocol.

Our simulations show that the stability and instability regions for the first moments practically coincide with those in panel 2.a of  Fig.~\ref{Toymodel}.

\begin{figure}[ht!]
    \centering
    \includegraphics[width=1\linewidth]{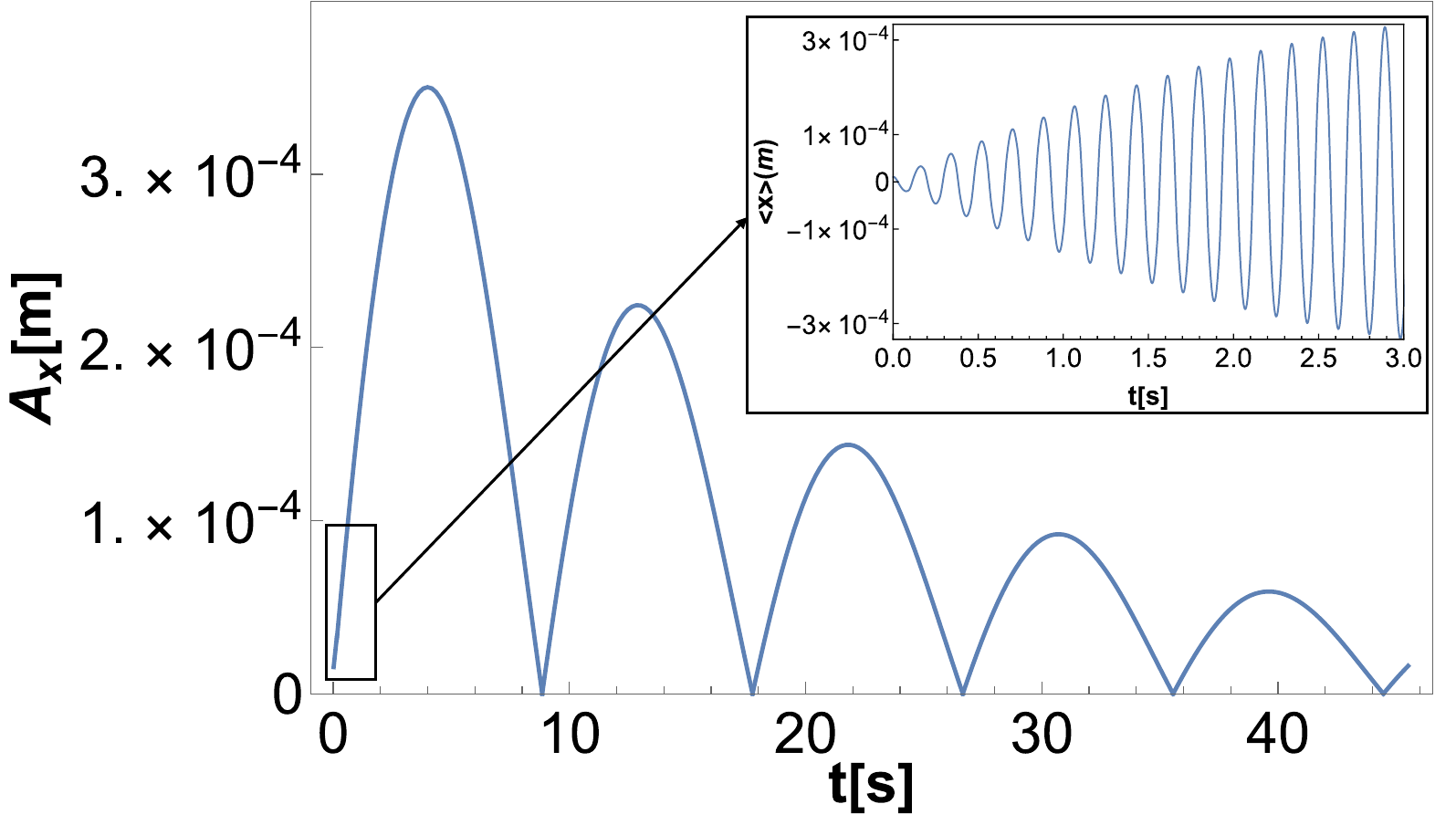}
    \caption{Time evolution of the envelope curve of the maximum amplitude of the oscillations for the position mean $\ave{\hat{x}}$ {for a stable solution}. The amplitude diminishes over time, and thus the system remains trapped. The parameters are fixed as in Tab.~\ref{RealParameters}, with initial values $\langle x(0)\rangle=10^{-5}\, \mathrm{m}$ and $\langle p(0)\rangle=10^{-9}\, \mathrm{kg\,m\,s^{-1}}$.}
    \label{First Moments Amplitude}
\end{figure}
\begin{figure}[h!]
    \centering
    \includegraphics[width=1\linewidth]{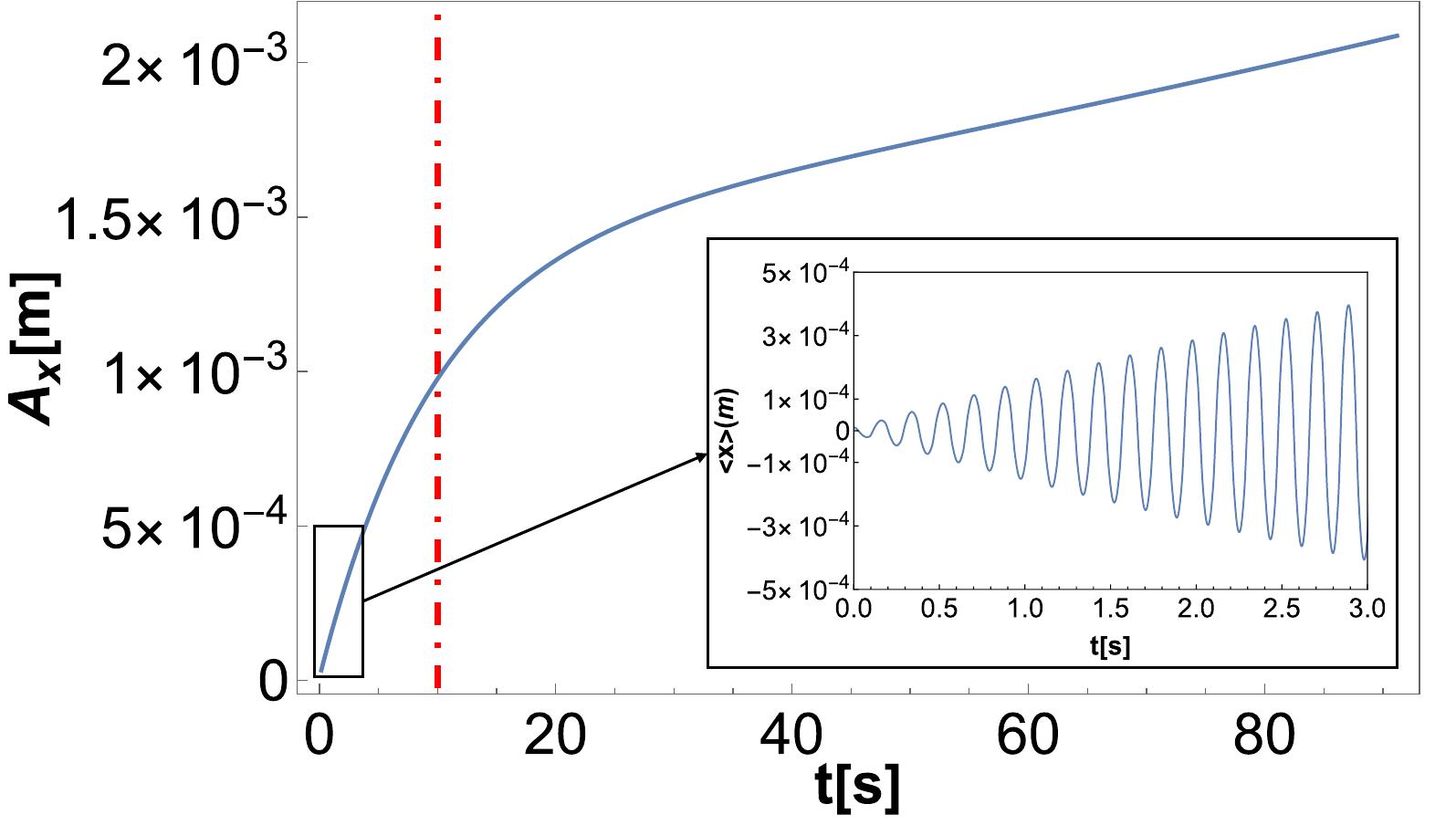}
    \caption{Time evolution of the envelope curve of the maximum amplitude of the oscillations for the position mean $\ave{\hat{x}}$ {for an unstable solution}. The amplitude increases over time, and the particle eventually leaves the trap. Parameters and initial values are as in Fig.~\ref{First Moments Amplitude}.}
    \label{FirstMomentsAmplitudeUnstable}
\end{figure}

\section{Time evolution of the second moments}

Let us consider as a starting point the Heisenberg equations of motion [cf. Eq.~\eqref{Langevin}] for the position $\hat{x}_t$ and momentum  $\hat{p}_t$ operators. Notice that the although the SN dynamics is nonlinear in the state $\psi_t(x)$, one can establish an effective Heisenberg picture \cite{Yang2013, helou2017measurable} and also include damping effects.

We analyze the dynamical evolution of the average second moments of the system, given by $V_{\text{ab}}=\frac{1}{2}\mathbb{E}[\langle\{\hat{a},\hat{b}\}\rangle_t]-\mathbb{E}[\ave{\hat{a}}]\mathbb{E}[\ave{\hat{b}}]$. 
Straightforward calculations lead to the following equations

\begin{equation}
\begin{split}
\dot{V}_{\text{xx}}&\!=\!\frac{2}{M}V_{\text{xp}}, \\
\dot{V}_{\text{xp}}&\!=\!\frac{1}{M}V_{\text{pp}}\!-\!M \omega_t^2 V_{\text{xx}}\! -\! M \omega_{\text{\tiny{SN}}}^2 (\mathbb{E}[\ave{\hat{x}^2}]\!-\!\mathbb{E}[\ave{\hat{x}}^2])\!-\!\gamma_\text{m}V_{\text{xp}},   \\
\dot{V}_{\text{pp}}&\!=\!-2 M \omega_t^2 V_{\text{xp}}\!-\!2 M \omega_{\text{\tiny{SN}}}^2 \left(\frac{1}{2}\mathbb{E}[\ave{ \{\hat{x},\hat{p} \} }]\!-\!\mathbb{E}[\ave{\hat{x}}\ave{\hat{p}}] \right)\\
&-2 \gamma_\text{m}V_{\text{pp}}\!+\!2 M \gamma_{\text{m}}k_{\text{B}}T. \\
\end{split}    
\end{equation}
and by defining the quantities 
\begin{equation} \label{Deviations}
\begin{split}
F_{\text{xp},t}&=\mathbb{E}[\ave{\hat{x}}^2]-(\mathbb{E}[\ave{\hat{x}}])^2, \\
F_{\text{pp},t}&=\mathbb{E}[\ave{\hat{x}}\ave{\hat{p}}]-\mathbb{E}[\ave{\hat{x}}]\mathbb{E}[\ave{\hat{p}}] 
\end{split}
\end{equation}
we can rewrite the dynamical evolution as 
\begin{equation} \label{SecMomEv}
\begin{split}
\dot{V}_{\text{xx}}&\!=\!\frac{2}{M}V_{\text{xp}}\\
\dot{V}_{\text{xp}}&\!=\!\frac{1}{M}V_{\text{pp}}\!-\!M \omega_{\text{q},t}^2 V_{\text{xx}}\!+\! M \omega_{\text{\tiny{SN}}}^2F_{\text{xp},t}\\
\dot{V}_{\text{pp}}&\!=\!-2 M \omega_{\text{q},t}^2 V_{\text{xp}}\!-\!2 \gamma_\text{m}V_{\text{pp}}\!+\!2M \gamma_\text{m}k_\text{B}T\!+\!2 M \omega_{\text{\tiny{SN}}}^2 F_{\text{pp},t},
\end{split}
\end{equation}
so by defining the vector $x_t$ as $x_t^T=(V_{\text{xx}} \quad V_{\text{xp}} \quad V_{\text{xp}})$, we can rewrite the dynamical evolution in the matrix form of Eq.~\eqref{Flononhom}.
From the definition of $\omega_t$ in Eq.~\eqref{squarewave}, in each of the two intervals defining each cycle of the evolution, the frequency $\omega_t$ takes a constant value. 
Without loss of generality, let us consider the evolution in the interval $0 < t < t_1$. In order to solve for the average values of $\hat{x}_t$ and $\hat{p}_t$ in Eq.~\eqref{Langevin}, we can use the Laplace transform, finding that
\begin{eqnarray}
\ave{\hat{x}}=G_{1,t} \langle \hat{x} \rangle_0 + G_{2,t} \langle \hat{p} \rangle_0 + \int_0^t \dd \tau F_{\text{th},\tau}G_{2,t-\tau},   
\end{eqnarray}
and
\begin{eqnarray}
\ave{\hat{p}}=\dot{G}_{1,t} \langle \hat{x} \rangle_0 + \dot{G}_{2,t} \langle \hat{p} \rangle_0 + \int_0^t \dd \tau F_{\text{th},\tau}\dot{G}_{2,t-\tau},
\end{eqnarray}
where
\begin{equation}
\begin{split}
G_{1,t}&=\frac{1}{2 \sqrt{\gamma_\text{m}^2-4 \omega^2}}\left(\gamma_\text{m}(e^{t F_+}-e^{tF_-}) \right.\\
&\left.+\sqrt{\gamma_\text{m}^2-4 \omega^2}(e^{t F_+}+e^{t F_-})\right),\\   
G_{2,t}&=\frac{1}{M \sqrt{\gamma_\text{m}-4 \omega^2}}(e^{t F_+}-e^{t F_-}),
\end{split}
\end{equation}
with
\begin{equation}
\Gamma_{\pm}=-\frac{\gamma_\text{m}}{2} \pm \frac{1}{2} \sqrt{\gamma_\text{m}^2 - 4 \omega^2}.   
\end{equation}
From these results, one can show that
\begin{equation}
\begin{split}
\mathbb{E}[\ave{\hat{x}}]&=G_1(t)\langle \hat{x} \rangle_0 + G_2(t) \langle \hat{p} \rangle_0,   \\
\mathbb{E}[\ave{\hat{x}}^2]&=(\mathbb{E}[\ave{\hat{x}}])^2+ 2 M \gamma_\text{m}k_\text{B} T \int_0^t \dd \tau G_{2,t-\tau}^2,
\end{split}
\end{equation}
and similarly
\begin{equation}
\begin{split}
\mathbb{E}[\ave{\hat{p}}]&=\dot{G}_1(t) \langle \hat{x} \rangle_0 + \dot{G}_2(t) \langle \hat{p} \rangle_0, \\
\mathbb{E}[\ave{\hat{p}}^2]&=(\mathbb{E}[\ave{\hat{p}}])^2+2 M \gamma_\text{m}k_\text{B}T \int_0^t \dd \tau \dot{G}_{2,t-\tau}^2.
\end{split}
\end{equation}
In addition:
\begin{equation}
\begin{split}
\mathbb{E}[\ave{\hat{x}}\ave{\hat{p}}]&=\mathbb{E}[\ave{\hat{x}}]\mathbb{E}[\ave{\hat{p}}]\\
&+2 M \gamma_\text{m}k_\text{B}T \int_0^t \dd \tau G_{2,t-\tau}\dot{G}_{2,t-\tau}.
\end{split}
\end{equation}
For our choice of the parameters, it turns out that $F_{\text{xp},t}$ and $F_{\text{pp},t}$ give a negligible contribution to the dynamics in Eq.~\eqref{SecMomEv}.

Neglecting these additional non-periodic terms, the stability of the second moments can be characterized analogously to what described in Appendix \ref{AppFirstMoments}. The analytic expressions of the eigenvalues are excessively lengthy and offer limited insight, even when $\gamma_m = 0$. Therefore, they are not presented here.

\end{document}